\RequirePackage{fix-cm}
\documentclass[twocolumn,final]{svjour3}          
\smartqed  
\usepackage{graphicx}
\journalname{Experimental Astronomy - EXPA-D-14-00036R3}
\graphicspath{{./}}
\begin{document}

\title{A Digital-Receiver for the Murchison Widefield Array
}


\def\RRI{$^{1}$}
\def\NRAO{$^{2}$}   
\def\CASS{$^{3}$} 
\def\ANU{$^{4}$}    
\def\MITK{$^{5}$}   
\def\ICRAR{$^{6}$}  
\def\CSIROCI{$^{7}$} 
\def\SWINBU{$^{8}$}  
\def\CFA{$^{9}$}     
\def\ASU{$^{10}$}    
\def\MITH{$^{11}$}   
\def\USYD{$^{12}$}   
\def\UWASS{$^{13}$}   
\def\VICUW{$^{14}$}   
\def\UWM{$^{15}$}     
\def\NCRA{$^{16}$}    
\def\UMEL{$^{17}$}    
\def\UTAS{$^{18}$}    
\def\SKAS{$^{19}$}   
\def\RHU{$^{20}$}     
\def\CSIROSS{$^{21}$} 
\def\UMICH{$^{22}$}   
\def\SKAUK{$^{23}$}    
\author{{Thiagaraj Prabu}\RRI\thanks{e-mail: \texttt{prabu@rri.res.in}}\and{K. S. Srivani}\RRI\and{D. Anish Roshi}\NRAO\and{P. A. Kamini}\RRI\and{S. Madhavi}\RRI\and
{David Emrich}\ICRAR\and{Brian Crosse}\ICRAR\and{Andrew J. Williams}\ICRAR\and{Mark Waterson}\ICRAR$^,$\ANU$^,$\SKAUK\and
{Avinash A. Deshpande}\RRI\and{N. Udaya Shankar}\RRI\and{Ravi Subrahmanyan}\RRI$^,$\CASS\and{Frank H. Briggs}\CASS$^,$\ANU\and
{Robert F. Goeke}\MITK\and{Steven J. Tingay}\CASS$^,$\ICRAR\and{Melanie Johnston-Hollitt}\VICUW\and 
{Gopalakrishna M R}\RRI\and
{Edward H. Morgan}\MITK\and{Joseph Pathikulangara}\CSIROCI\and{John D. Bunton}\CSIROSS\and{Grant Hampson}\CSIROSS\and
%
{Christopher Williams}\MITK\and{Stephen M. Ord}\ICRAR\and{Randall B. Wayth}\CASS$^,$\ICRAR\and{Deepak Kumar}\RRI\and
%
{Miguel F. Morales}\UWASS\and{Ludi deSouza}\CSIROSS$^,$\USYD\and{Eric Kratzenberg}\MITH\and{D. Pallot}\ICRAR\and{Russell McWhirter}\MITH\and
{Bryna J. Hazelton}\UWASS\and{Wayne Arcus}\ICRAR\and{David G. Barnes}\SWINBU\and {Gianni Bernardi}\CFA$^,$\SKAS$^,$\RHU\and{T. Booler}\ICRAR\and
{Judd D. Bowman}\ASU\and{Roger J. Cappallo}\MITH\and{Brian E. Corey}\MITH\and{Lincoln J. Greenhill}\CFA\and
{David Herne}\ICRAR\and{Jacqueline N. Hewitt}\MITK\and{David L. Kaplan}\UWM\and 
{Justin C. Kasper}\UMICH$^,$\CFA\and{Barton B. Kincaid}\MITH\and{Ronald Koenig}\CSIROSS\and
{Colin J. Lonsdale}\MITH\and
{Mervyn J. Lynch}\ICRAR\and 
{Daniel A. Mitchell}\CASS$^,$\CSIROSS\and{Divya Oberoi}\MITH$^,$\NCRA\and{Ronald A. Remillard}\MITK\and{Alan E. Rogers}\MITH\and
{Joseph E. Salah}\MITH\and{Robert J. Sault}\UMEL\and{Jamie B. Stevens}\UTAS$^,$\CSIROSS\and{S. E. Tremblay}\CASS$^,$\ICRAR\and
{Rachel L. Webster}\CASS$^,$\UMEL\and{Alan R. Whitney}\MITH\and{Stuart B. Wyithe}\CASS$^,$\UMEL\and
}
\authorrunning{MWA Collaboration}
\institute{$^{1}$ Raman Research Institute (RRI), Bangalore, India.
$^{2}$ National Radio Astronomy Observatory, Green Bank, USA.
$^{3}$ ARC Centre of Excellence for All-sky Astrophysics {(CAASTRO)}.
$^{4}$ Australian National University (ANU), Canberra, Australia.
$^{5}$ MIT Kavli Institute, Boston, USA.
$^{6}$ International Centre for Radio Astronomy Research - Curtin University, Perth, Australia.
$^{7}$ CSIRO Computational Informatics, Australia.
$^{8}$ Swinburne University of Technology, Melbourne, Australia.
$^{9}$ Harvard-Smithsonian Center for Astrophysics, Cambridge, USA.
$^{10}$ Arizona State University, Tempe, USA.
$^{11}$ MIT Haystack Observatory, Westford, USA.  
$^{12}$ University of Sydney, Sydney, Australia.  
$^{13}$ University of Washington-Seattle, Seattle, USA.
$^{14}$ Victoria University of Wellington, New Zealand.
$^{15}$ University of Wisconsin-Milwaukee, Milwaukee, USA.
$^{16}$ National Centre for Radio Astrophysics - TIFR, Pune, India.
$^{17}$ University of Melbourne, Melbourne, Australia.
$^{18}$ University of Tasmania, Hobart, Australia.
$^{19}$ Square Kilometre Array South Africa (SKA SA), Cape Town, South Africa.
$^{20}$ Department of Physics and Electronics, Rhodes University, Grahamstown, South Africa. 
$^{21}$ CSIRO Astronomy and Space Science, Australia.
$^{22}$ University of Michigan, Ann Arbor, USA.
$^{23}$ SKA Organisation, Jodrell Bank Observatory, UK.
}

\date{Received  / Accepted }


\maketitle
\begin{abstract}
An FPGA-based digital-receiver has been developed for a low-frequency imaging radio interferometer, the Murchison Widefield Array (MWA). The MWA, located at the Murchison Radio-astronomy Observatory (MRO) in Western Australia, consists of 128 dual-polarized aperture-array elements (tiles) operating between 80 and 300\,MHz, with a total processed bandwidth of 30.72 MHz for each polarization. Radio-frequency signals from the tiles are amplified and band limited using analog signal conditioning units; sampled and channelized by digital-receivers. The signals from eight tiles are processed by a single digital-receiver, thus requiring 16 digital-receivers for the MWA. The main function of the digital-receivers is to digitize the broad-band signals from each tile, channelize them to form the sky-band, and transport it through optical fibers to a centrally located correlator for further processing. The digital-receiver firmware also implements functions to measure the signal power, perform power equalization across the band, detect interference-like events, and invoke diagnostic modes. The digital-receiver is controlled by high-level programs running on a single-board-computer. This paper presents the digital-receiver design, implementation, current status, and plans for future enhancements.
\end{abstract}



\section{Introduction}    
\label{intro}
\noindent An FPGA-based digital processing module that we refer to as the digital-receiver in this paper is especially developed for a new generation, wide field-of-view, low-frequency radio telescope: the Murchison Widefield Array (MWA) \cite{LONS}\cite{TING}\cite{BOWM}. The MWA is an international collaboration between Australian, New Zealand, US, and Indian institutions and the telescope is located within the radio-quiet Murchison Radio-astronomy Observatory (MRO) in  Western Australia, one of the two Square Kilometre Array (SKA) designated sites. After having demonstrated a prototype telescope consisting of 32 aperture array elements (tiles), a larger telescope made of 128 tiles, each operating between 80 and 300\,MHz with a total usable processed bandwidth (sky-band) of 30.72\,MHz has now been realized. Specialized subsystems, such as the digital-receiver described in this paper have been developed for this  telescope. 

The primary role of the digital-receiver is to sample \mbox{80-300}\,MHz analog streams from the dual-polarized tiles of the MWA, channelize to select the sky-band, and send it downstream for further processing. 

The design, integration and operation of digital-receivers into the MWA has addressed many engineering challenges associated with a remote and high temperature telescope site, many of which are applicable to the design and construction of future instruments to be deployed on the MRO and possibly for the SKA. 

The non-availability of any ready-to-use solutions for the digital processing led us to the development of a dedicated digital-receiver for the MWA. Crucial considerations such as the need to simultaneously sample and channelize a large band from many inputs, perform fine-grain parallel operations in processing, portability of the hardware at the field, required mains power for processing, programmability, flexibility to incorporate multiple operating modes, provisions for future upgrades in the design, and finally the availability of critical resources within the collaboration led us to develop an FPGA-based digital-receiver for the MWA. We had first implemented an early version of the digital-receiver for the prototype telescope, and now have implemented the full MWA digital-receiver as explained in this paper.

Listed below are aspects from the digital-receiver design that appear of interest to future systems. 
The digital-receiver design that we describe in this paper is evolved around the architecture of MWA, where:
\begin{itemize}
    \item a cluster of tiles being dealt-with in one processing unit,
    \item the processing units are optimally located in the field,
    \item the sampling clocks for the digitizers are locally generated at each processing unit using a distributed reference, 
    \item a synchronized operation with all other processing units is achieved using a distributed reference, 
    \item the data produced at the processing unit is transmitted to a remote central processing station using optical fibers, and   
    \item the processing units are sufficiently shielded in the field to prevent interference.
\end{itemize} 

In this paper, as an introduction we present the signal flow in the MWA, followed by an outline of major considerations in the design. Here, we discuss the crucial choices that shaped the digital-receiver architecture. Then we explain the signal processing sections of the digital-receiver, operating modes and diagnostic features. Then we present the details of the implemented hardware. In the last section, we review the importance of this work from the context of the upcoming SKA, and conclude highlighting the results obtained from using the digital-receiver with the MWA.

\section{Design Philosophy} 

Since the MWA is realized as a precursor to a major telescope of the future, the design of the subsystems acquired a specific scrutiny. The design philosophy adapted in the digital-receiver design is centered around an early work by Briggs \cite{BRIGS}. 



At the MWA telescope site, instead of locating the receiver electronics in a central facility they are distributed across the array so that digital processing can be performed early in the signal path to minimize the delays and degradation of signal. To minimize the total number of stations at the telescope site, signals from a groups of tiles are processed in common stations. A group size of 8 tiles per processing station was chosen based on an optimal configuration of the digitizers and the tile distribution pattern expected for the array.

A spectral measurement study at the telescope site reported in Bowman et al. \cite{BOWM2} supported designing the system to use 8-bit digitization preceded by analog band pre-selection filtering and signal conditioning. Another development work [Section \S4] by Bunton et al. at CSIRO helped us to use a baseband channelizer configuration for the digital-receiver where the full RF band from each tile is handled by a dual analog-to-digital converter (ADC) followed by a Virtex-4 FPGA-based polyphase filter bank for channelization.  

A choice of about 30\,MHz for the instantaneous processed band was considered adequate based on data distribution cost, complexity, real-time processing capability, and the needed continuum sensitivity. During the implementation of the digital-receiver, the following choices were made: a) to sample the 80-300\,MHz bands using a sampling clock at a higher rate of 655.36\,MHz, so that the sampled band is free from aliasing as well as being divisible in binary steps, and b) to channelize the sampled band using a 512-point FFT based polyphase filter bank. The filter bank output consists of 256 equally-sized coarse channels covering the full RF band from DC to 327.68\,MHz. These choices fixed the amount of instantaneous sky-band that is transmitted to the central processing station for real-time processing at 30.7\,MHz wide, formed out of 24 coarse channels of each 1.28\,MHz width. Considering the expected dynamic range of the input signal and that gain equalization will be applied to the sky-band data stream prior to requantization, a choice was made to represent them using a \mbox{5-bit} real, \mbox{5-bit} imaginary pair format, reducing the data transmission requirements.  

The digital-receivers are distributed across a wide area in the telescope site and need to maintain coherent sampling and synchronized channelization across all stations. For this purpose a frequency reference for sampling and a GPS-based time reference for synchronization are distributed from a central station over a fiber-optic link, and the receivers include circuitry to synchronize their local clocks to these two reference signals. 

The digital-receivers are typically operated as remote stations, and hence a robust control and monitoring capability was needed. For this purpose, a distributed monitoring-and-control (M\&C) system was designed. At the central facility a scheduling, monitoring and control software system coordinates operation of all the receivers over Ethernet connections to the remote receivers. Each of the stations includes a single-board computer running an embedded Linux operating system to bridge the digital-receivers with the remote M\&C facility and to control and configure the programmable hardware. This platform allows collection of comprehensive internal sensor data as well as calculating and supplying a variety of meta-data information to the central processing station. 

The telescope site at the MRO experiences a very wide temperature variation between the days and nights, and between the summer and winter months. For this purpose a weather-proof, RF shielded enclosure was designed to house the analog processing electronics, digital-receivers and the single-board-computer [Section \S2.2]. The amount of power dissipated by the electronics and the need to provide a stable temperature environment resulted in a requirement for active cooling of the enclosure, and the most cost-effective solution for this was found to be a mains-powered commercial refrigeration unit. This subsystem is controlled by the single-board computer using internal temperature sensors located at several places in the electronics package. Locating any digital hardware within a sensitive array such as the MWA requires provisions to arrest any radio-frequency-interference (RFI) that they may tend to emit. The MWA receiver unit that houses the digital-receiver also houses the sensitive analog signal conditioner, and therefore the receiver unit is specifically designed to contain the digital-receiver and the analog signal conditioner in suitably shielded modules to control RFI emission and susceptibility.


\subsection{System Overview}

\begin{figure}
\begin{center}  
  \includegraphics[width=0.5\textwidth]{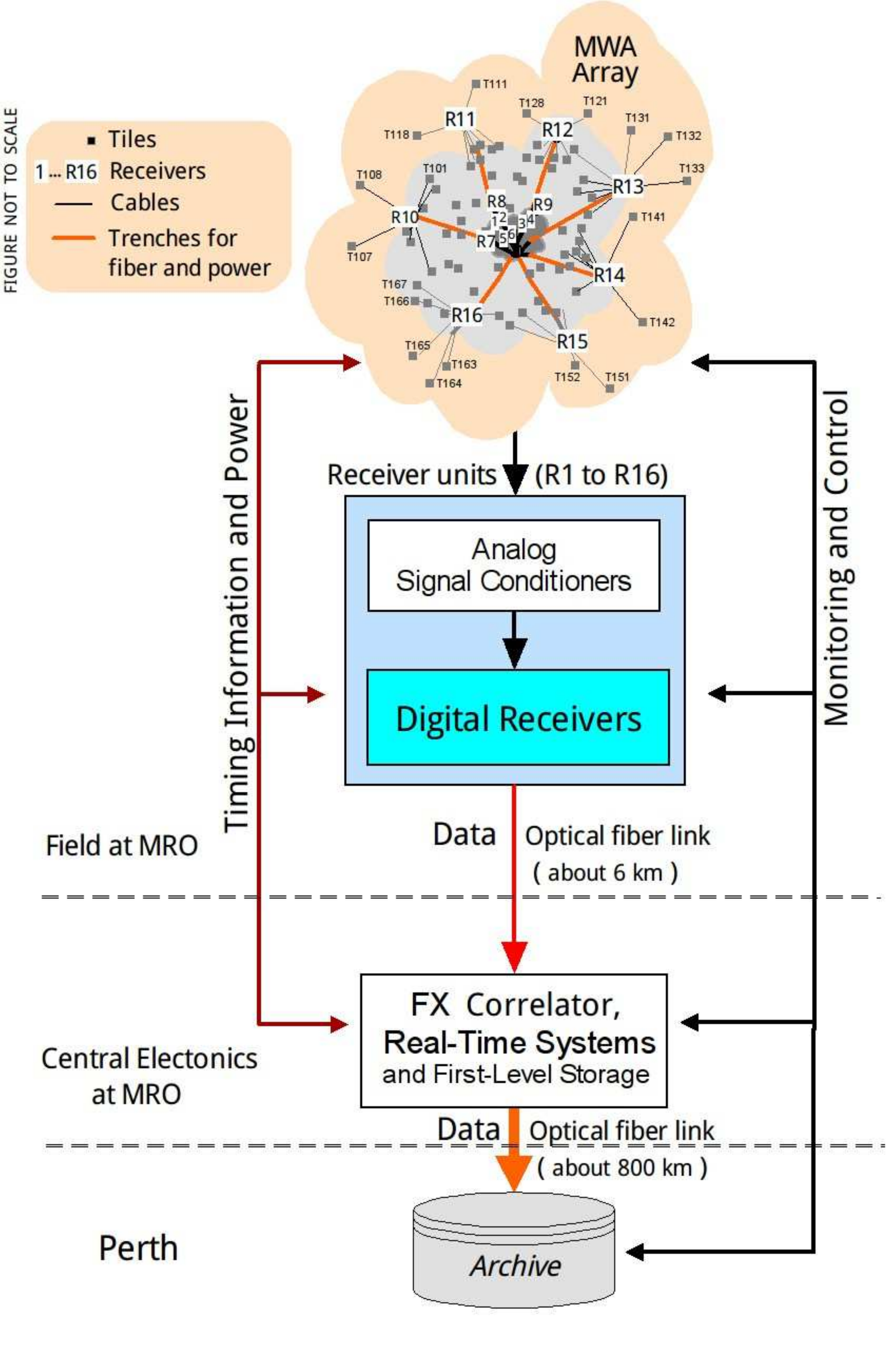}  
\end{center}  
\caption{An overview of the subsystems and their connectivity at the MWA.}
\label{fig:1}       
\end{figure}

A signal flow diagram of the main subsystems in the MWA is shown in \mbox{Figure}\,1.
The MWA tiles are arrays of 16 dual-polarized, crossed dipole antennas arranged in a regular 4x4 pattern and distributed over a 1.5\,km radius area with a distribution optimized for the main observing modes of the telescope \cite{BEAR}. Each of the MWA tiles is equipped with a digitally controlled, analog time-delay beamformer providing a primary pointing capability to each tile. After beamforming, a pair of analog signals corresponding to the two polarizations is transmitted through coaxial cables (with an optional whitening-filter) up to a maximum length of 500\,m to a nearby station where the MWA receiver units are located. 
 
The receiver unit contains the electronics needed for the analog and digital processing of signals for eight dual-polarized tiles. Because each unit handles only eight tiles, a total of 16 such MWA receiver units are deployed in the telescope site for 128 tiles. Inside the  MWA receiver unit, the analog processing is provided by the analog signal conditioner (ASC), and the digital processing by the digital-receivers. 

The tile signals that arrive at the receiver unit are first amplified and band-limited (80 to 300\,MHz) in the ASC. This subsystem provides impedance conversion, amplification (see AMPLIFIER in Fig. 2) and band-limiting filters (see FILTER in Fig.2). Total amplification for each signal chain is adjustable via step attenuators over a 60\,dB range with a resolution of 1\,dB, giving a maximum amplification of +33\,dB at 130\,MHz. A band-limited signal-band with a 3\,dB band-pass between 80-300\,MHz and a stop-band attenuation better than 30\,dB at 327\,MHz is passed on to the digital-receiver. Inside each digital-receiver, the signals from eight tiles are continuously sampled, channelized to form the 30.72 MHz wide sky-band, and the data is transmitted out on optical fibers.  

An underground cable network is used on the telescope site to distribute electric power to the receivers and optical fibers for carrying the data, timing signals, and information for control and monitoring of the receivers. The sky-band data travels over a 6\,km distance to reach the central electronics facility, where the FX-correlators\,\cite{CHIK-CORR} are located. At this stage, the sky-band from all tiles are further channelized into 10\,KHz wide sub-bands and passed on for real-time correlation. The visibilities \cite{MORAN} thus obtained are moved to a first-level temporary storage for access by any real-time systems, and then are transported on optical fibers over 600 km to the archival facility\,\cite{DAVE} at Perth, meant for long-term storage and access by the users.

\subsection{MWA Receiver hardware}  

\begin {figure*}
\begin{center}
\includegraphics[width=1\textwidth]{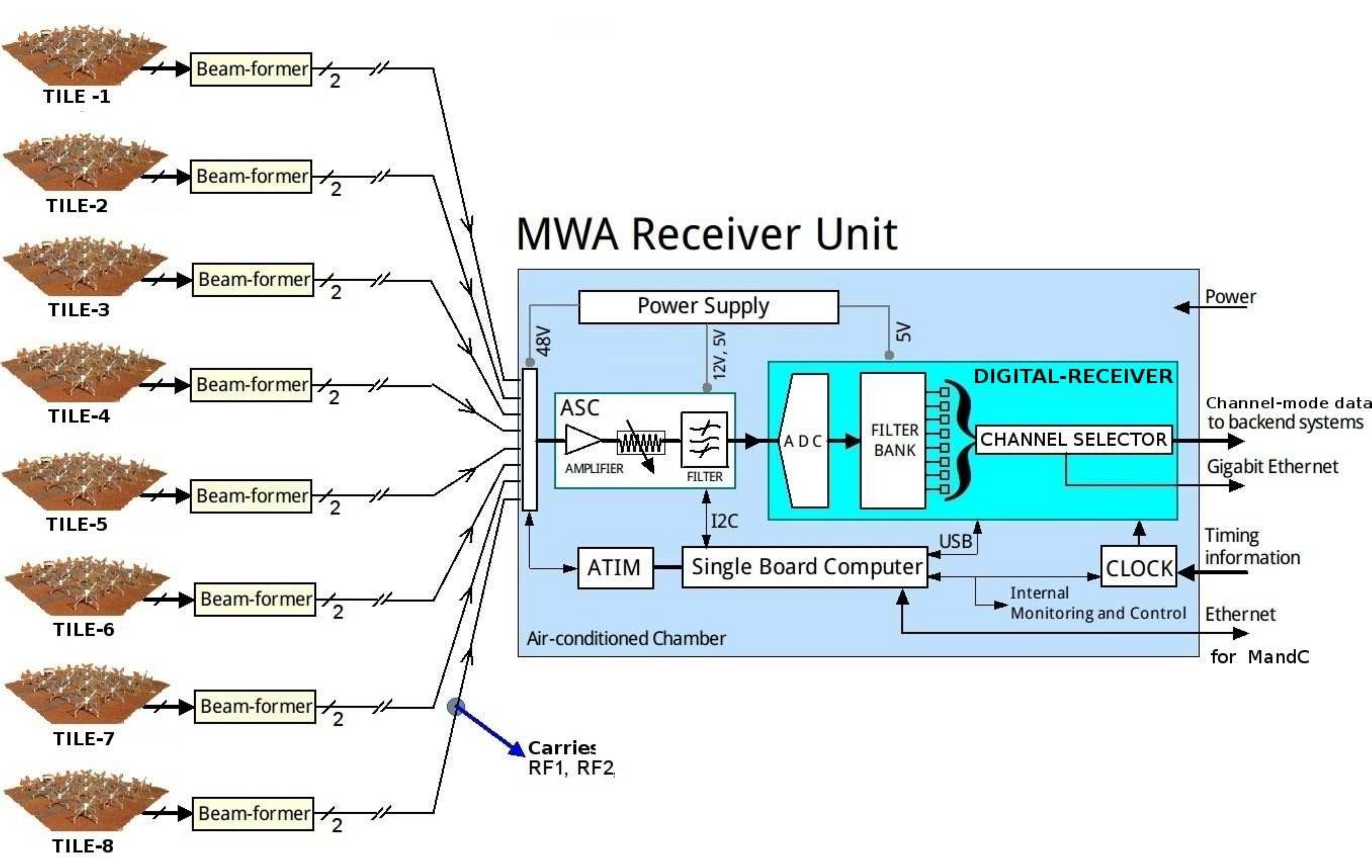}
\end{center}
\caption{A schematic view of the MWA receiver unit along with its subsystems. The MWA receiver contains the electronics needed for the analog and digital processing of signals for eight dual-polarized tiles. The analog processing is provided by the analog signal conditioner (ASC), and the digital processing is provided by the digital-receiver. The MWA receiver unit is housed in an air conditioned chamber which also provides the electro-magnetic shielding needed for operating sensitive instruments in the field. A total of 16 such MWA receiver units are deployed in the MWA site.}
\label{aba:fig2}
\end {figure*}

At the MWA telescope site, the digital-receiver along with other crucial subsystems are housed inside the MWA receiver unit. A schematic view of the MWA receiver unit and its external connectivity is shown in Figure 2. 
Each of the  MWA receiver unit consists of several sub-components: 

\begin{itemize}
 \item  Analog signal conditioner (ASC):
  \begin{itemize}
    \item to equalize the signal power received from the tiles
    \item to filter a bandpass between 80 to 300\,MHz
  \end{itemize}    
  \item Antenna interface module (ATIM) to control the tile pointing
  \item Digital-receiver for sampling (ADC) and channelization 
  \item Clock-module to generate:
  \begin{itemize}
    \item sampling clock at 655.36\,MHz
    \item processing clock at 163.84\,MHz
    \item synchronization pulse at intervals of one second
  \end{itemize}  
  \item  Single-board-computer (SBC) for M\&C interface
  \item  Regulated power supply for the internal sub-components and the tile beamformers
  \item  Air-conditioning system to maintain a stable operating temperature
  \item  Metal enclosure providing an EMI shield.
\end{itemize}

\noindent The MWA receiver unit's external interfaces include:

\begin{itemize}

 \item Eight pairs of coaxial cables:
  \begin{itemize}
    \item to bring the two polarization signals from each tile  
    \item to send phasing commands to the beamformers
    \item to convey DC power to the beamformer and tile low-noise amplifiers 
    \item to monitor the tile and beamformer status.
  \end{itemize} 
   \item Three optical fibers to transmit the channelized data to the correlator
   \item One optical fiber to deliver:
  \begin{itemize}
    \item Reference for the sampling clock
    \item Synchronization information based on a 1\,s modulation
   \end {itemize}
 \item One optical fiber pair to provide the M\&C interface through Ethernet
 \item One optical fiber pair to provide the Gigabit Ethernet output
 \item Mains (line-voltage) power:
  \begin {itemize}
    \item to feed all internal power-supplies and the air-conditioning unit. 
  \end {itemize}
\end{itemize} 


\section{Design of the Digital-Receiver}

\subsection{Overview of the Architecture}
\begin {figure*}
\begin{center}
\includegraphics[width=1.05\textwidth]{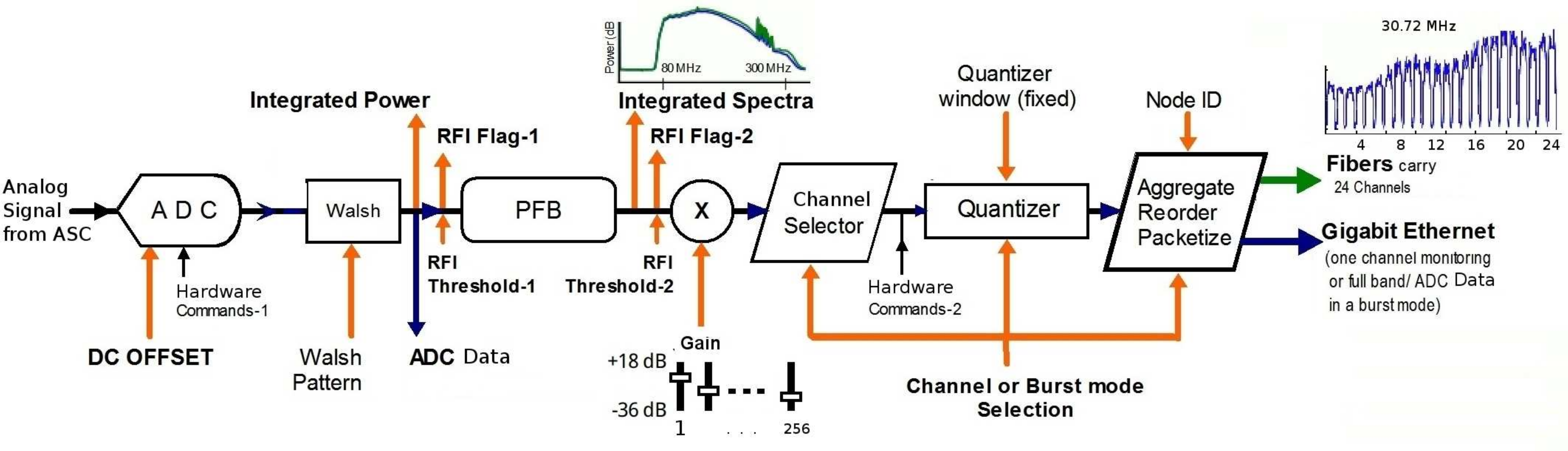}
\end{center}
\caption{An overview of the signal processing architecture in the digital-receiver.}
\label{aba:fig3}
\end {figure*}
 
An outline of the architecture developed for the digital-receiver in its primary operating mode is presented in this section. The digital-receiver contains the processing modules essential for sampling and channelizing the analog signal streams from eight dual-polarized tiles. An overview of this design is shown in Figure 3.

Inside each digital-receiver, there are 16 identical processing pipelines (to process signals from eight dual-polarized tiles) and a common aggregation logic. Each of these pipelines consists of ADC, Walsh module, polyphase filter bank (PFB), gain module, channel selection module, and a requantizer module. 

The ADCs operate on a 655.36\,MHz sampling clock and digitize the input band into an eight-bit wide data streams. The ADC outputs are passed on through a Walsh demodulator section \cite{THOM} into the PFB module. The PFB section operates on 4096 data samples and channelizes the 327.68\,MHz wide input band into 256 coarse channels including a DC component. The coarse channels are read out of the PFB as \mbox{16-bit} real, \mbox{16-bit} imaginary pairs (16+16) while the redundant conjugate part of spectrum is discarded because of the input being a real sequence.
The coarse channels from the PFB enter the channel selection module, where each coarse channel is multiplied by a gain of between -36\,dB to +18\,dB to equalize power across the band and to achieve a preferred bit-occupancy at the requantizer output. A selection of 24 arbitrarily arranged or contiguous coarse channels are then chosen to form the instantaneous sky-band. This data is further quantized to \mbox{5-bit} real, \mbox{5-bit} imaginary pairs (5+5). 
  
The selected channels are sent out continuously at a rate of 1.28\,MHz from each of the 16-pipelines to the aggregator. The aggregator gathers the data, performs reordering, formatting and transmission of the channels on optical fibers to the real-time systems located at the central electronics facility.  

For special purpose applications and diagnostics, it is also possible to configure the channel-selection module to select all 256 channels, but pass them in a burst form [Section \S3.3.2] to the aggregator module. The burst form outputs are tapped through a Gigabit Ethernet port available in the aggregator module. 
 
The digital-receiver also continuously measures the signal power at the ADC outputs, computes average spectra for monitoring purposes, and can detect interference-like events. These features and a set of diagnostic modes to facilitate testing of the digital-receivers in the field, are governed by the M\&C system through the  M\&C interface [Section \S3.5.3] and dedicated high-level programs running on the single-board computer. 

The digital-receiver uses optical fibers, Gigabit Ethernet, and USB ports for transmitting different forms of data. Section  \S3.3.1 presents the use of optical fibers in the channel mode. The burst and raw modes, described in Section \S3.3.2, transmit data through the Gigabit Ethernet port. The meta-data products described in Section \S3.3.3 are accessed by the M\&C system through the USB port. The signal processing logic in the digital-receiver is implemented using the FPGAs, and the details about their implementation is described in Section \S4. 
%
\subsection{Signal Processing}
\subsubsection{Input Quantization} 

The signal power at the ADC inputs needs to be set to an optimal level for efficient use of the digitizer dynamic range. The typical signal power at different inputs can vary due to beam position in the sky, the cable attenuation, and due to any RFI in the band. Any clipping or saturation of the signal due to strong RFI, such as caused by the satellite emissions, must be avoided before it is fed to the ADCs. For this purpose, a remote station can assess the signal power being fed to the ADCs by reading:
\begin{itemize}
   \item the ADC output voltage, using the Gigabit Ethernet port
   \item the ADC output power, using the M\&C interface
   \item the RFI event detection flag, using the M\&C interface
   \item the power spectra, using the M\&C interface. 
\end{itemize} 
The first option requires switching the digital-receiver to a special mode, to route the ADC output voltage to the Gigabit Ethernet port. The later three options are part of the meta-data functionality [Section \S3.4] and hence can be used while the digital-receiver is operating in the normal modes [Section \S3.3]. After assessing the signal power, the amount of amplification at the ASC can be adjusted by the M\&C so that an optimal signal power is seen by the ADCs.

\subsubsection{Walsh Demodulation}
The Walsh module removes any phase-switching performed in the tile beamformer, and passes the data to PFB module for further processing. For this purpose, a sixteen state demodulation sequence unique for each of the polarizations and tiles will switch at intervals of 625\,$\mu$s (a period corresponding to 100 sets of 4096 ADC samples) per state, with the full sequence of 16 steps repeating once every 10\,ms. The demodulation is carried out by flipping the polarity of the digitized voltage data streams being fed from the ADCs to the PFBs. The switching signal used for demodulation is  made available to synchronize the external phase modulators.

\subsubsection{Polyphase Filter Bank} 

The MWA digital-receiver uses a critically sampled polyphase filter bank network to implement channelization \cite{BELL-POLY}\cite{VAID}. 


The PFB splits the 327.68\,MHz wide sampled band into 256 coarse channels, with each channel being centered 1.28\,MHz away from the adjacent channel and 1.28\,MHz wide. The component filters of the PFB are 8-taps wide and feed 512-point FFTs, thus forming a 4096-tap FIR response for each sub-band defined by a coarse channel.

\begin {figure*}
\begin{center}
\includegraphics[width=0.5\textwidth]{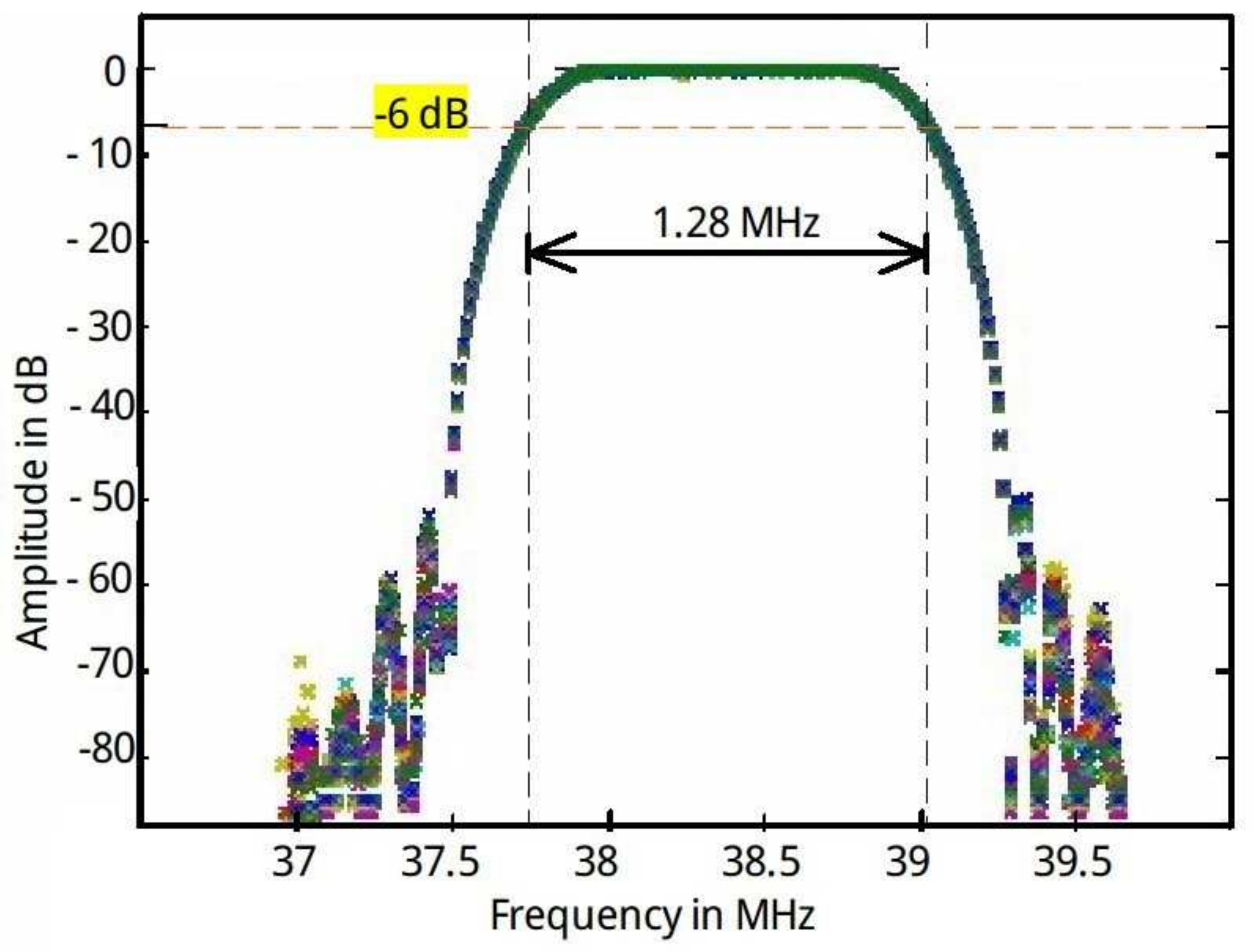}
\end{center}
\caption{ A laboratory measurement made of the PFB response for coarse channel\,30. This measurement is made by sweeping a CW RF signal at the digital-receiver (ADC) input. The leakages from the adjacent channel side-lobes are seen below -50\,dB level, and the adjacent channels overlap at -6\,dB level.}
\label{aba:fig8}
\end {figure*}

A measurement made on the PFB-response is shown in \mbox{Figure\,4}. The plot shown is an overlay of 16 independent measurements made while a CW RF signal at the ADC input was swept across the frequencies corresponding to coarse channel 30 (37 to 40\,MHz). The digital-receiver was configured in burst mode [Section \S3.3.2] to make these measurements. The measurements show the leakage from the prominent side-lobes into the adjacent bands are below -50\,dB, and the adjacent bands overlap at the -6\,dB level. 

\subsubsection{Requantizer}
The number of bits representing the complex output is quantized to \mbox{5-bit} real, \mbox{5-bit} imaginary pairs for the channel mode [Section \S3.3.1] of the digital-receiver. The requantizer uses symmetric round-off \cite{ROUND-OFF} and saturation logic while resampling numbers to represent them in a lower-bit representation. In this bit reduction implementation, the voltage gain section available at the output of the PFB is used to equalize the pass-band gain variations and maintain the bit-occupancy needed to optimally represent the radio astronomy noise signal in the quantized outputs with a headroom reserved for interference. A gain correction function can be obtained from the power spectrum by taking its square root and normalizing to a peak value.


\subsection{Modes of Operation}

The digital-receiver can be configured to operate in: a) Channel mode, b) Burst mode, c) Raw mode and d) Diagnostic Mode. The channel mode is specifically developed for the science operations with the telescope. The burst, raw and diagnostic modes are developed for the monitoring and commissioning operations. 

\subsubsection{Channel mode}

In channel mode, the digital-receiver is programmed to continuously output a specific set of 24 channels needed for an observation. In this mode, the aggregator gathers the selected 24 coarse channels for both polarizations of the eight tiles and reorders them to make it convenient for further processing and distribution at the correlator multipliers. The reordered channels are formatted into packets [Section \S3.5.1] and transmitted out using three optical fibers, with each fiber carrying eight coarse channels from all eight dual-polarized tiles. The packets consist of a header portion to identify the packet, receiver, fiber, a time-stamp, a checksum and a data portion containing eight coarse channels from eight dual-polarized tiles. The time-stamp field is used to align data from different-receivers before correlation, and is created from a count based on the GPS one-second tick at each digital-receiver. Any one of the 24 coarse channel voltages, for all eight tiles, being transmitted from the digital-receiver can also be continuously monitored using the Gigabit Ethernet port available in the aggregator [Section \S3.5.2].

\subsubsection{Burst Mode}

The digital-receiver can also be programmed to operate in a burst mode, in which all 256 coarse channels of each PFB for all eight tiles are collected once every 1024 PFB frames and sent out in bursts. The burst mode outputs are available through the Gigabit Ethernet port. In this mode, the native 16+16 bit representation of the PFB outputs are preserved, but the full band is only available once every 799.7\,${\mu}$s. Burst mode allows to simultaneously examine the entire band (DC to 327.68\,MHz) in full precision, by trading off continuous time-series output for bandwidth.

\subsubsection{Raw Mode}
The raw mode is operationally similar to the burst mode, however in this mode 256 time samples from each ADC, capturing both polarizations of all eight tiles, are collected once every 799.7\,${\mu}$s. The outputs retain the native number representation of the ADCs and are available through the Gigabit Ethernet port.  

\subsubsection{Diagnostic Modes} 
The diagnostic modes are incorporated in the digital-receivers to aid testing them in the field during maintenance. By switching a digital-receiver into the diagnostic mode, test patterns can be injected at strategic locations in the signal path including: the output of ADC, the output of the PFB, the input of the aggregator, and at the input of the optical fibers, thus helping to systematically test and isolate faulted sections in the field. Several special-purpose registers are also incorporated in the digital-receiver to monitor critical signals such as the synchronizing pulse and clock. In addition, laboratory based test modes were also developed to test and qualify a new hardware before it is commissioned in the field.

\subsection{Meta-Data}
 During a normal operation, the digital-receiver generates a set of supplementary information:
\begin{itemize}
  \item ADC output power, measured and integrated over a second 
  \item RFI event detection flag at each ADC output (obtained by counting the number of times the integrated power from each ADC exceeds a programmable-threshold) 
  \item Power spectra for each of the eight tiles (obtained by squaring and averaging the 16+16 bit complex outputs of the PFB and integrating over one second) 
  \item RFI event detection flag for all 256 coarse channels (estimated by counting the number of times each coarse channel power exceeds a programmable-threshold). 
\end{itemize}

This supplementary information set forms the primary signal meta-data from the digital-receiver. The meta-data are refreshed at one second intervals and can be periodically monitored through the USB port. The meta-data products are available when the digital-receiver is configured to operate in channel, burst or raw mode.

\subsection{Data Interfaces} 

\subsubsection{Optical Fiber data links}
The fiber ports in the aggregator transmit a minimally-formated data from the receiver to the central processing station. The data is assembled into packets (conforming to a gt\_custom protocol from Xilinx) by the aggregator and tagged with a sequence number to provide limited error detection and facilitate re-assembly of the data sequence.  Section  \S3.3.1 presents the use of optical fibers in the channel mode. 

\subsubsection{Gigabit Ethernet Port}
The aggregator also implements a standard Gigabit Ethernet port to provide a simplified means of monitoring the digital-receiver during commissioning and for tests during maintenance, as well as to transmit the high-throughput data stream for the burst-mode [Section \S3.3.2 \& \S3.3.3]. This port allows direct access to: a) the digital-receiver outputs that are normally accessible only after processing by the correlator, and b) the data streams that are internal to the digital-receiver and before they are reduced in the channelizer. A dedicated fiber port is incorporated in the digital-receiver for routing the Gigabit Ethernet outputs to the central electronics facility. For this purpose, the data from the digital-receiver are formatted to comply with the user datagram protocol (UDP) and sent out as Internet packets (UDP/IP) on a conventional computer network. 

\subsubsection{USB Port} 
Since the digital-receiver is a programmable instrument, a flexible interface is required to send commands to the individual digital-receiver boards and monitor their operation. For this purpose, the FPGAs in the digital-receiver [Section \S4] interface to the single-board computer through USB ports. Using the USB interface, tile specific control, configuration and monitoring information can be routed to/from specific channelizers in the digital-receiver from/to the M\&C. The meta-data products [Section \S3.4] are also accessed by the M\&C system through the same USB port.


\subsection{Overview of the Operation} 

A typical initialization sequence required to bring up the digital-receivers after each power cycle is shown in Figure 5. This process is automatically executed by the receiver SBC under the control of the central M\&C processes, so that the initialization of the digital-receiver firmware can be coordinated across the entire telescope. There are three stages involved in this process: 1) Loading the FPGA firmware (Booting), 2) Configuring the programmable registers inside the FPGAs, and 3) Selecting a specific mode for the operation. During the first stage, the FPGAs (a total of nine) [Section \S4] in the digital receiver are loaded with the serial images containing their firmware programs. A slave serial mode FPGA configuration interface is used for this purpose. The firmware programs are stored on solid-state disk in the receiver SBC and can be replaced with updated images in the field by downloading the new code to the SBC. In the second stage, the registers inside the FPGAs are configured with identifiers unique to each digital-receiver, and empirically-determined offsets and initial gain coefficients for each signal chain. During the third stage of initialization, the FPGAs are configured to function for a specific observing mode and to select the instantaneous band required for the observation.

\begin {figure}
\begin{center}
\includegraphics[width=0.5\textwidth]{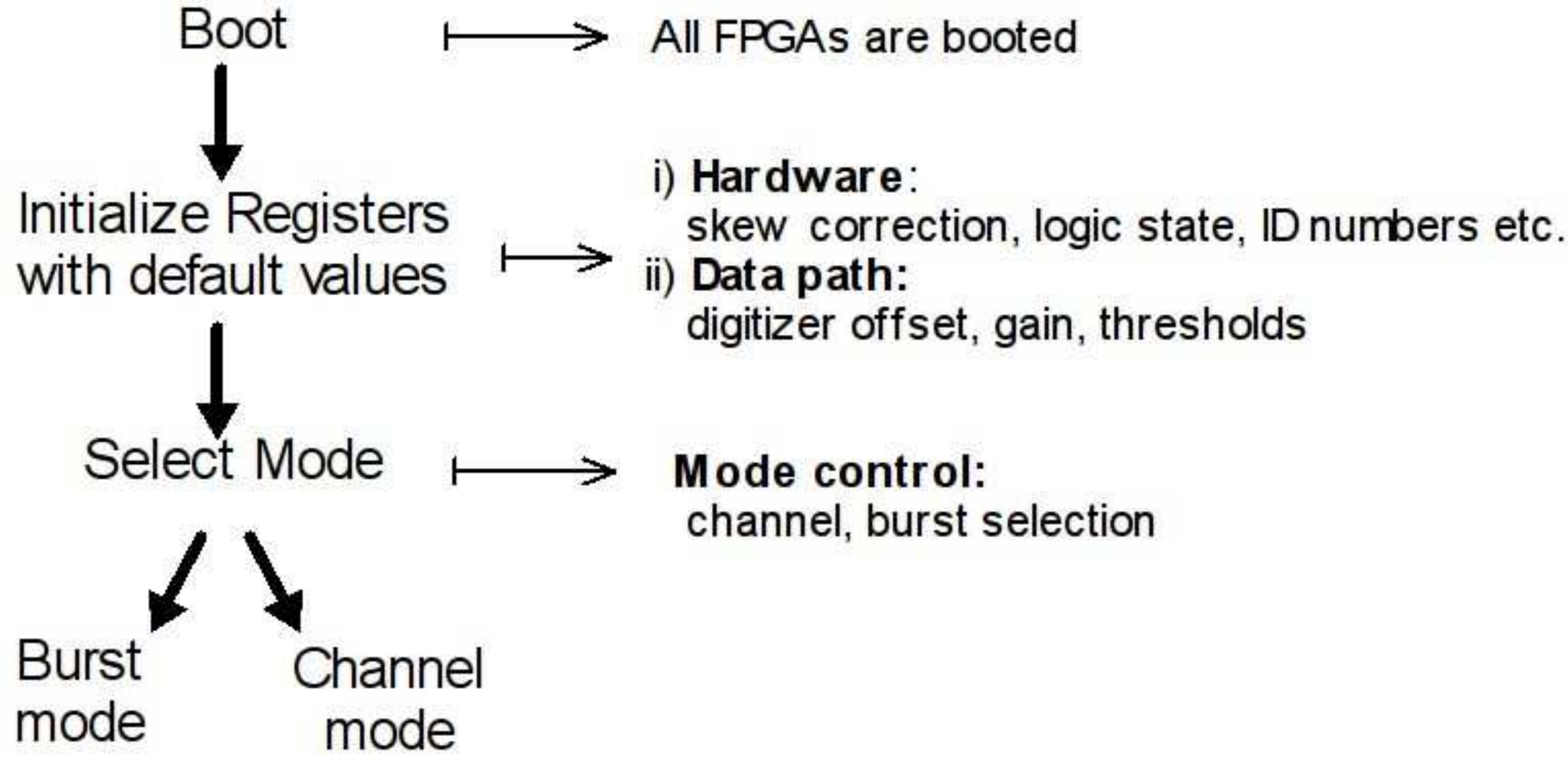}
\end{center}
\caption{Typical operation of the digital-receiver involve booting the FPGAs, programming the internal registers and switching to a desired mode.}
\label{aba:fig5}
\end {figure}


\section{Implementation}

 
\subsection{Hardware Implementation}  
The hardware required for 16-pipelines (to process signals from eight dual-polarized tiles) of signal processing and a common aggregator at each digital-receiver is implemented using two kinds of FPGA boards. A schematic view of this implementation is shown in Figure 6.

The main signal processing at each pipeline consists of analog-to-digital conversion and channelization. The analog-to-digital converters (ADC) used in the digital-receiver are e2v\textsuperscript{\textregistered} AT84AD001C giga-sample digitizers developed by ATMEL\textsuperscript{\textregistered}. Each of these chips house two ADCs, and there are eight such e2v\textsuperscript{\textregistered} chips used in each digital-receiver. The signal processing required for the channelizer is carried out using eight Xilinx\textsuperscript{\textregistered} Virtex-4 SX35\footnotemark[1] FPGAs with each FPGA housing two polyphase filter banks. This section of the hardware is implemented in two identical analog-to-digital cum filter-bank (ADFB\footnotemark[2]) circuit boards, with each board having four ADCs and four FPGAs.

\footnotetext[1]{XCV4SX35-10FFG668I or C}

The common aggregator-formatter (AgFo) circuit board consists of logic for gathering data from 16-inputs, reordering, packaging, high-speed serializing and the modules for electrical-to-optical fiber conversion. The logic portion of aggregator is implemented in a Xilinx\textsuperscript{\textregistered} Virtex-5 SX50T\footnotemark[3] FPGA. 

\footnotetext[3]{XC5VSX50T-1FFG665C}

The electrical-to-optical conversion modules are implemented using FINISAR\textsuperscript{\textregistered}  FTLF-1421 SFP modules. The entire aggregator hardware is implemented in a single circuit board (AgFo), which contains the FPGA and four SFP modules. Out of these SFP modules, the transmit sections of three of the SFP transceiver units are used to route the channel mode data to three optical fibers in an 8b/10b encoded format, and the entire fourth SFP module is used to route the Gigabit Ethernet. The receive section of one of the SFP is used to implement a fiber loop-back mode for diagnostics, while the remaining two SFPs receive sections are not used.

The ADFBs and the AgFo boards are integrated using a custom designed back-plane board. This back-plane carries the PFB outputs to the aggregator, distributes: clock, synchronization signals and power, and provides a test-port for diagnostic data capture. 

\footnotetext[2]{\it {The ADFB board, PFB library,  FTDI card and the associated low-level programs were developed by the ICT division of CSIRO.}}

\begin {figure*}
\begin{center}
\includegraphics[width=1\textwidth]{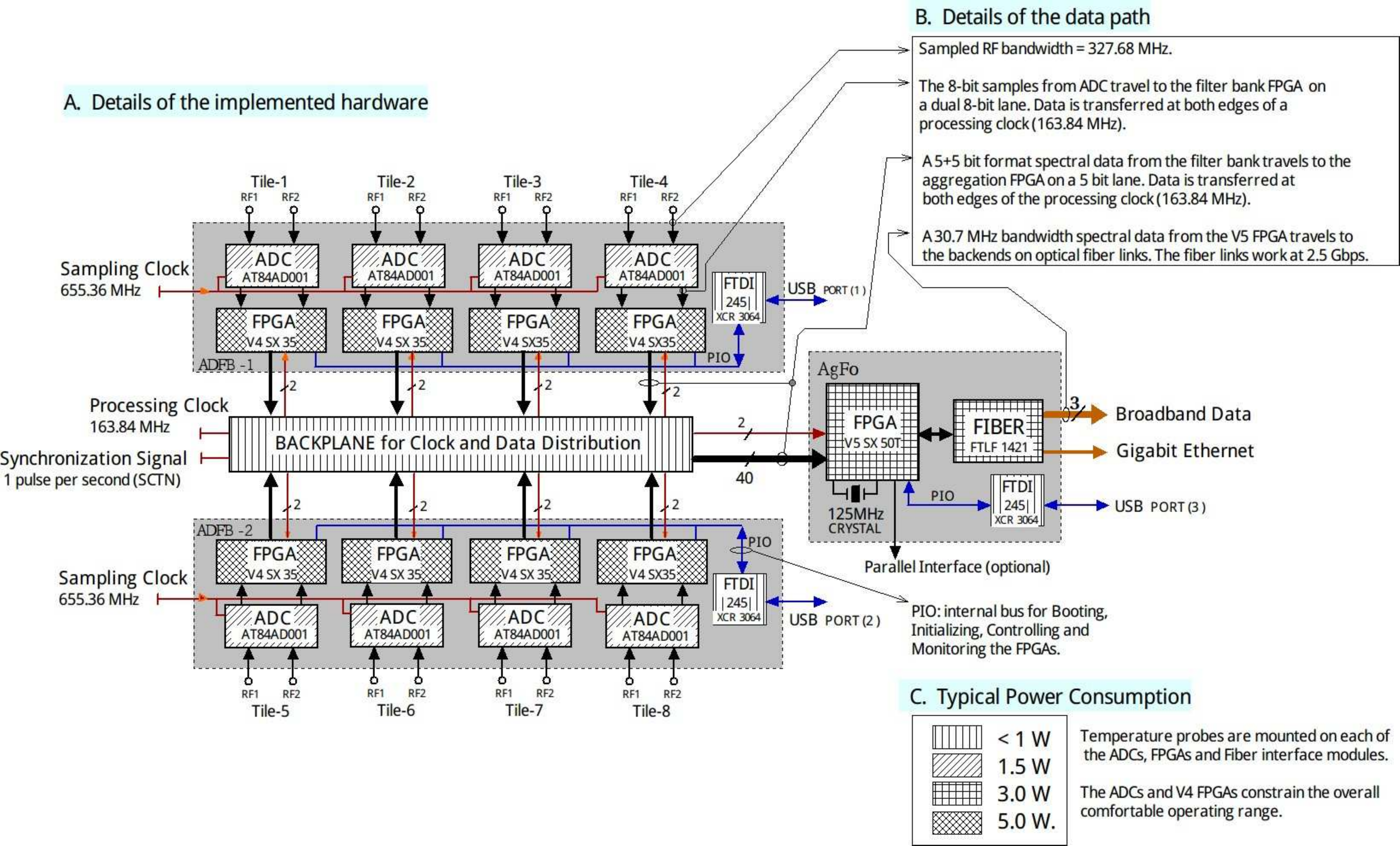}
\end{center}
\caption{An overview of the digital-receiver hardware implementaion.}
\label{aba:fig4}
\end {figure*}
The Gigabit Ethernet over fiber-optic is implemented baesd on a Xilinx IP library. The USB interface is implemented based on an FTDI\textsuperscript{\textregistered} FIFO chip and a CPLD. The FIFO chip interfaces to the USB bus, and a firmware logic in the CPLD interfaces to the FPGA using a parallel input/output (PIO) protocol. A set of dedicated control interface programs were developed in C-language to provide a high-level interface to the digital-receiver for the M\&C operation.

The Walsh switching system is implemented by means of a switchable 180 degree phase shifter in the beamformer and a digital inversion step in the digital-receiver firmware. The pattern of beamformer switching across the receiver tile set is controlled by a fixed state table embedded in the antenna-tile control (ATIM) and synchronized with a matching demodulation table in the digital-receiver firmware. In the present operating modes of the digital-receiver, this feature has not been activated in the field so far.

\subsubsection{Timing Signals}  
The digital-receiver operation is governed by: 1) the sampling clock to the ADCs, 2) the processing clock to the FPGAs, 3) the synchronizing pulse, and 4) the reference clock for the data transmission over fibers. The sampling clock, processing clock and the synchronizing pulse are supplied to the digital-receiver boards from a clock-module located in each receiver-unit (see Figure 2). 

The clock-module receives its input over a uni-directional fiber link from the central electronics facility at MRO. At the central electronics facility, a reference synthesizer produces a clock at 163.84\,MHz frequency which is modulated by a GPS-derived 1 pulse-per-second tick. This signal is split and amplified to directly drive the transmit sections of 16 SFP fiber links to distribute the clock to all 16 MWA receiver-units located in the field.

A clock-module in each receiver-unit demodulates this signal and produces a phase-locked analog clock at 655.36\,MHz, a digital clock at 163.84\,MHz, and a synchronizing pulse having a width of 6.1\,ns at 1\,s periodicity. The 655.36\,MHz clock feeds the ADC modules of the digital-receiver for coherent sampling. The samples from the ADC are processed in the FPGAs using a parallel architecture, where 163.84\,MHz clock is sufficient for the processing to keep up with the ADC output streams. 
Synchronization of sampling epochs between the 16 digital-receivers and across the 16 pipelines within a digital-receiver is achieved by the synchronizing pulse. The processing clock and the synchronizing pulse are distributed to all FPGAs through the digital-receiver backplane. Upon power on, the synchronizing pulse generation logic in the clock-module is reset with the incoming GPS-derived 1\,s tick; after that initialization, consecutive synchronizing-pulses are derived solely by dividing the 163.84\,MHz clock at the clock-module. The SBC can also monitor and reinitialize the clock-module by means of an interface port.

The digital-receivers transmit data at 2.5 Gbps through the optical fibers. For this purpose, a reference clock at 125 MHz is generated by a separate crystal oscillator in the AgFo board. This clock is free-running and is not synchronized with the sampling clock.

\subsubsection{Power Dissipation} 

The digital-receiver is operated from a regulated 5\,V\,DC power supply. The power consumption in the digital-receiver varies as a function of the operating mode, from about 25\,W in standby condition to 75\,W in channel mode operation. The major power consumption in the digital receiver comes from the FPGAs (4 to 5\.W per FPGA). The ADCs and the fiber interface components also consume significant power of about 1.5\,W per ADC and 1\,W per fiber module. The FPGAs are fitted with metal heat-sinks and fans to quickly dissipate the heat, and temperature sensors are mounted on these critical components to monitor and send feedback to the air-conditioning system. 

\subsection{Firmware Implementation}
The logic for signal processing pipelines and data aggregation are implemented in the \mbox{FPGAs} as firmware programs. All modules of ADFB, except for the PFB, were designed using VHDL for the Virtex-4 FPGAs. The aggregator logic is designed in VHDL for the Virtex-5 FPGA. A combination of Xilinx synthesis software ISE 9.2.04 through ISE\,12.04 and Modelsim\,5.8b were used to compile and simulate the designs. 

The PFB module is incorporated as an EDIF library provided by CSIRO. A simplified block diagram of the PFB as seen by the firmware modules is given in Figure 7. Major resources used for implementing one PFB module in Virtex-4 SX35\footnotemark[1] is listed in Table-1.

\begin{table}[ht]
\caption{Resources used by the PFB in FPGA Virtex-4 XCV4SX35-10FFG668} 
\centering 
\begin{tabular} {c c c} 
\hline
Resource Name & Units used  & Percentage used   \\ [0.5ex] 
\hline
      &       &     \\
DSP48 & 68    & 35\% \\
RAM16 & 43    & 22\% \\
SLICE & 3413  & 22\% \\
[1ex] 
\hline  
\end{tabular}
\label{table:nonlin} 
\end{table}

\begin {figure*}
\begin{center}
\includegraphics[width=1\textwidth]{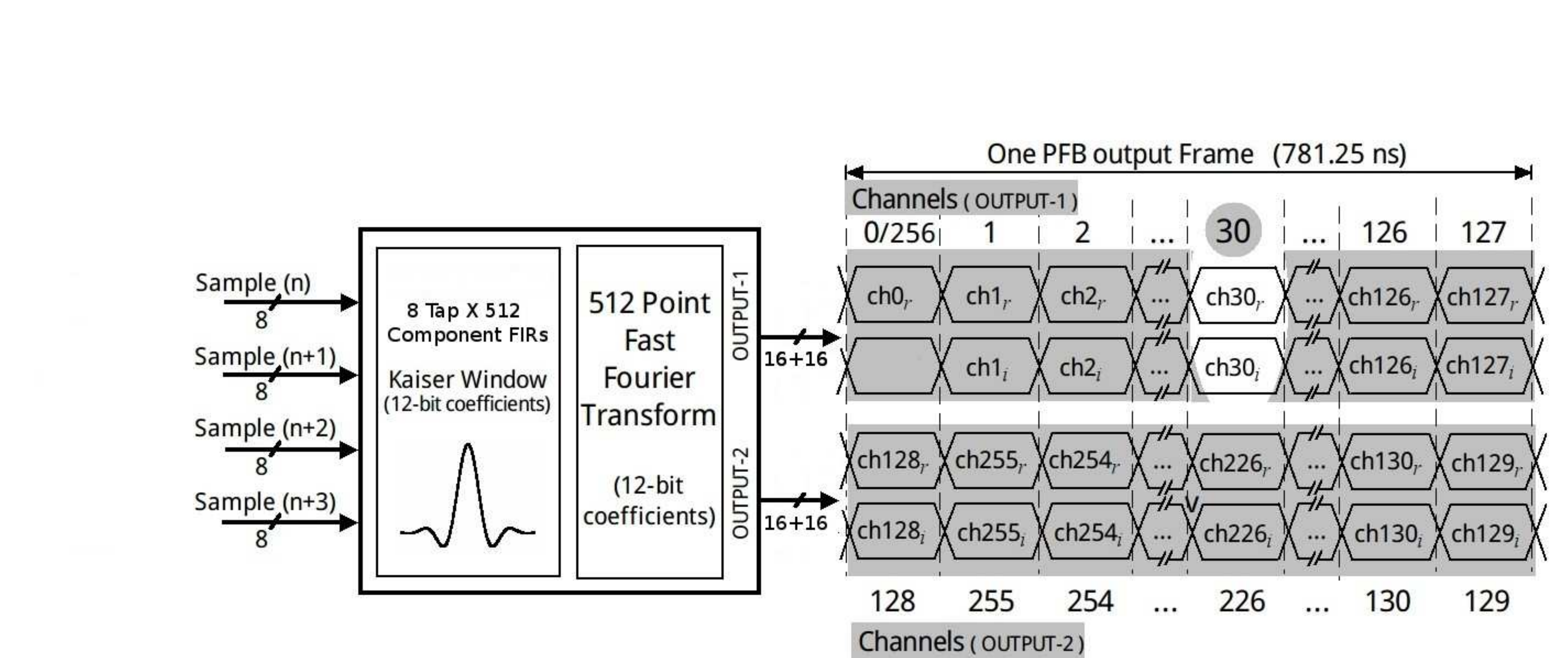}
\end{center}
\caption{ An overview of the PFB module as seen by the FPGA firmware. The inputs to the PFB are fed in a sequence formed by four consecutive time samples, and the outputs from the PFB appear with a specific channel order along the two-sequences output-1 and 2 that appear simultaneously with a specific mapping between the data-sequence and the channel number as seen in the figure. The PFB outputs are \mbox{16-bit} real (ch\textit{N}{r}), \mbox{16-bit} imaginary (ch\textit{N}{i}) pair complex numbers, where \textit{N} being the channel number between 0 to 255.}
\label{aba:fig6}
\end {figure*}
The input samples to the PFB are 9-bits wide, represented in a two's complement format and divided into four sequences. The samples from ADC, after they pass through the Walsh-module, are sign-extended to 9-bits and fed to the PFB. 
The filter coefficients and twiddle-factors used in the PFB are 12-bits wide. The PFB outputs are \mbox{16-bit} real, \mbox{16-bit} imaginary pair complex numbers. The PFB outputs appear simultaneously in two sequences (shown as outputs 1 and 2, in Figure 7), with a unique mapping between the output sequence and the channel-number. 


\section{Future Upgrades} 

Many interesting astrophysical phenomena associated with the broadband fast-transients can be investigated using a time-domain data. For this purpose, the digital-receiver described in this paper is being enhanced to provide a wide-band (80-300\,MHz) voltage-beam-forming mode. In this new mode, the signals from eight tiles serviced by the digital-receiver to be coherently phased to produce a higher gain voltage-beam with a narrow angular response over the entire MWA band. For this purpose, the required delay and phase corrections are applied at the sampler and channelizer outputs respectively.
 This new mode is accomplished through hardware and firmware enhancements to pass the full 80-300\,MHz band from the PFB outputs to the aggregator FPGA, and by incorporating the beam-forming logic in the FPGA. The reduction of 16 data pipelines to only two data-streams allows the wide-band voltage-beam from each digital-receiver to be transmitted on the available three data fibers to the central electronics facility. The enhanced digital-receiver also maintains a complete downward compatibility with all its existing capabilities and operational modes \cite{PRAB}. 

\section{Summary and Conclusion}
After successfully testing the digital-receiver at the laboratory and field, 16 digital-receiver systems have been integrated with the 128 tiles of the Murchison Widefield Array at the Murchison Radio-astronomy Observatory in Western Australia. One unique challenge, in this complex design, was to make hardware system suitable for operation in the remote,radio quiet and high-temperature telescope site at the MRO \cite{TING}. The choice of using a single-board computer for the active-cooling control, helped to apply needed control algorithms and successfully operate the receiver electronics over a wide ambient air temperature range at the MRO.

The architecture used in the design of the front-end digital processing in MWA alos acquires a significance for the future systems, because the MWA is a precursor to the future SKA. 

The MWA digital-receiver employs high-speed analog-to-digital converters for sampling a 327\,MHz band, and an FPGA-based hardware to implement signal processing tasks that are inherently deterministic and exhibit fine-grain parallelism. Considering the technology advancements in the performance of FPGA and ADC devices, the philosophy used in the current design presents a valuable consideration for designing digital processing to sample and process much larger bandwidths for the future large telescopes \cite{BUNT}.     

The front-end digital instrumentation that is to be designed for the future large telescopes is required to provide excellent support for the synthesis imaging back-ends and a phased array mode with multi-beaming capabilities. The MWA digital-receivers have been designed to support both these desired aspects and therefore appear as an appropriate reference for the design of the digital front-end for future telescopes.

A filter bank is an important functional part in digital processing, and in the MWA digital-receiver we have implemented a critically sampled PFB. This choice has provided useful feedback in exploring the critically sampled PFB functionality and about the resource utilization in the FPGAs.   

The instrumentation for the future large telescopes are also required to provide immense capability for control and monitoring operations from remote stations, flexible programmability to change configurations, capability to detect RFI like events in real-time, ability to support remote commissioning tests and built-in diagnostic capabilities to aid maintenance. The MWA digital-receiver has been designed to contain these capabilities and hence it is a good starting reference for future systems.

And finally, the digital-receiver is also designed as a drop-in module, with connectivity to the  MWA receiver unit through copper cable connectors and optical fiber sockets; it allows an easy upgrade of the entire digital module as the technology advances in future.

The success of the digital-receiver instrument for the Murchison Widefield Array can be seen by a number of science results produced during the digital-receiver integration period utilizing the MWA prototype \cite{BERN} \cite{BELL} \cite{OBER}  \cite{MCKI-CENA} \cite{MCKI-MOON} \cite{TING-SOLAR} and further results achieved following full commissioning of the digital-receiver system with all 128 tiles \cite{BHAT} \cite{HIND}.

\begin{acknowledgements}

This scientific work makes use of the Murchison Radio-astronomy Observatory, operated by CSIRO. We acknowledge the Wajarri Yamatji people as the traditional owners of the Observatory site. Support for the MWA comes from the U.S. National Science Foundation (grants AST-0457585, PHY-0835713, CAREER-0847753, and AST-0908884), the Australian Research Council (LIEF grants {LE0775621} and LE0882938), the U.S. Air Force Office of Scientific Research (grant FA9550-0510247), and the Centre for All-sky Astrophysics (an Australian Research Council Centre of Excellence funded by grant CE110001020). Support is also provided by the Smithsonian Astrophysical Observatory, the MIT School of Science, the Raman Research Institute, the Australian National University, and the Victoria University of Wellington (via grant MED-E1799 from the New Zealand Ministry of Economic Development and an IBM Shared University Research Grant). The Australian Federal government provides additional support via the Commonwealth Scientific and Industrial Research Organisation (CSIRO), National Collaborative Research Infrastructure Strategy, Education Investment Fund, and the Australia India Strategic Research Fund, and Astronomy Australia Limited, under contract to Curtin University. We acknowledge the iVEC Petabyte Data Store, the Initiative in Innovative Computing and the CUDA Center for Excellence sponsored by NVIDIA at Harvard University, and the International Centre for Radio Astronomy Research (ICRAR), a Joint Venture of Curtin University and The University of Western Australia, funded by the Western Australian State government. The National Radio Astronomy Observatory (NRAO) is a facility of the National Science Foundation operated under cooperative agreement by Associated Universities, Inc. The authors also acknowledge the contributions from the Raman Research Institute members: Somashekar for the X1 expedition, Girish for testing the PFB, Kasturi for qualifying the ADFB analog inputs and for developing a noise source, Ananth, Arasi, Mamatha, Sandhya and Sujatha for making the 16 channel analog input system used for testing the digital-receiver, Vinutha for the work with the USB card, Sarabagopalan for the cold test setup, Vinod for the ADFB ID cards, Wenny for making the Clock/SCTN distribution board for the 32T digital-receivers, Peeyush Prasad and C\,R\,Subramanya for providing the UDP code for the AgFo, Mechanical Engineering Services for making the power-splitter enclosures and the heat-sink mounts, Ravi Sankar for the electro-plating work, Mamatha Bai and RRI Administration for taking care of the administrative logistics, purchase team for the procurements, and the computer group for their support with linux and networking; Franz Schlagenhaufer at ICRAR for the Electromagnetic Compatibility tests of the MWA receiver and Jonathan Tickner at ICRAR for the support during the receiver integration tests; Mark Leach and Ron Ekers at CSIRO for sharing their expertise; and the Halleens at Boolardy for the warm hospitality during the site visits.
\end{acknowledgements}


\begin{thebibliography}{}
%
%
\bibitem{LONS}
Lonsdale {\it et al.,} {\it ``The Murchison Widefield Array: Design Overview"}, Proceedings of the IEEE, Vol 97, Issue 8, Pages(s): 1497-1506  (2009) 
\bibitem{TING}
Tingay {\it et al.,} {\it ``The Murchison Widefield Array: The Square Kilometre Array Precursor at Low Radio Frequencies"}, Publications of the Astronomical Society of Australia, Volume 30, id.e007 21 pp (2013)
\bibitem{BOWM}
Bowman {\it et al.,} {\it ``Science With the Murchison Widefield Array"}, Publications of the Astronomical Society of Australia Volume 30, id.e031 28 pp (2013)
\bibitem{BRIGS}
Briggs, H. F., {\it ``MWA-LFD Receiver Node Subsystem"}, MWA knowledge tree, (2007)
\bibitem{BOWM2}
Bowman et al., {\it ``Field Deployment of Prototype Antenna Tiles for the Mileura Widefield Array Low Frequency Demonstrator"}  The Astronomical Journal, 133:1505Y1518  (April 2007)
\bibitem{BEAR} 
Beardsley et al., {\it ``A new layout optimization technique for interferometric arrays, applied to the MWA"}, Monthly Notices of the Royal Astronomical Society, Volume 425, Issue 3, Pages(s): 1781-1788 (2012)
\bibitem{CHIK-CORR}
Chikada, Y., et al. {\it ``A very fast FFT spectrum analyzer for radio astronomy"} Acoustics, Speech, and Signal Processing, IEEE International Conference on ICASSP '86, Page9s): 2907-2910 Vol-11 (1987)
\bibitem{MORAN}
Thompson, A.R., Moran, J.M., Swenson, G.W. (Jr), {\it ``Interferometry and Synthesis in Radio Astronomy"} 2nd Edition, ISBN: 978-0-471-25492-8, (May 2001)
\bibitem{DAVE}
Wu, C., Wicenec, A., Pallot, D., Checcucci, C.,  {\it ``Optimising NGAS for the MWA Archive"} 
Experimental Astronomy
Volume 36, Issue 3, pp 679-694  (December 2013)
\bibitem{THOM}
Thompson, A.R., Clark, B.G.,Granlund, J., {\it ``An Application of Walsh Functions in Radio Astronomy Instrumentation"} Electromagnetic Compatibility, IEEE Transactions on Volume: EMC-20 , Issue: 3, Page(s): 451-453 (1978)
\bibitem{ROUND-OFF}
Schneeweiss, H., Komlos, J., Ahmad, A. S. 
{\t ``Symmetric and asymmetric rounding: a review and some new results "} Advances in Statistical Analysis,
Volume 94, Issue 3, pp 247-271, (September 2010)
\bibitem{BELL-POLY}
Bellanger, M.G., Bonnerot, G., Coudreuse, M., {\it ``Digital filtering by polyphase network: Application to sample-rate alteration and filter banks"}, Acoustics, Speech and Signal Processing, IEEE Transactions on Volume: 24, Issue: 2, DOI: 10.1109/TASSP.1976.1162788, Pages(s): 109-114 (1976)
\bibitem{VAID}
Vaidyanathan, P.P., {\it ``Multirate Digital Filters, Filter Banks, Polyphase Networks, and Applications: A Tutorial"}, Proceedings of the IEEE, Volume 78, Issue: 1, DOI:10.1109/5.52200,  Pages(s): 56-93 (1990)\par
\bibitem{PRAB}
Prabu et al., {\it ``A Full-band Voltage Beamforming mode for the Murchison Widefield Array Digital Receiver"}, MWSKY 2013 conference proceedings, ASI (submitted, 2014)
\bibitem{BUNT}
Bunton, J.D., {\it ``SKA Correlator Advances"}, Experimental Astronomy, Volume 17, Issue 1-3, Pages(s): 251-259 (2004)
\bibitem{BERN}
Bernardi et al., {\it ``A 189\,MHz, 2400 square degree polarization survey with the Murchison Widefield Array 32-element prototype"}, The Astrophysical Journal, Volume 771, Issue 2, article id. 105, 16 pp (2013)
\bibitem{BELL}
Bell et al., {\it ``A survey for transients and variables with the Murchison Widefield Array 32-tile prototype at 154\,MHz"},	Monthly Notices of the Royal Astronomical Society, Volume 438, Issue 1, Pages(s): 352-367 (2014)
\bibitem{OBER}
Oberoi et al., {\it ``First Spectroscopic Imaging Observations of the Sun at Low Radio Frequencies with the Murchison Widefield Array Prototype"}, The Astrophysical Journal Letters, Volume 728, Issue 2, article id. L27  (2011)
\bibitem{MCKI-CENA}
McKinley et al., {\it ``The giant lobes of Centaurus A observed at 118\,MHz with the Murchison Widefield Array"}, Monthly Notices of the Royal Astronomical Society, Volume 436, Issue 2, p.1286-1301 (2013)
\bibitem{MCKI-MOON}
McKinley et al., {\it ``Low Frequency Observations of the Moon with the Murchison Widefield Array"}, The Astronomical Journal, Volume 145, Issue 1, article id. 23, 9 pp (2013)
\bibitem{TING-SOLAR}
Tingay et al., {\it ``The Murchison Widefield Array: solar science with the low frequency SKA Percursor"}, Journal of Physics: Conference Series, Volume 440, Issue 1, article id. 012033 (2013)
\bibitem{BHAT}
Bhat et al., {\it ``The low-frequency characteristics of PSR J0437-4715 observed with the Murchison Widefield Array"}, The Astrophysical Journal Letters, Volume 791, Issue 2, article id. L32, 6 pp. (2014) 

\bibitem{HIND}
Hindson et al., {\it ``The First Murchison Widefield Array low frequency radio observations of cluster scale non-thermal emission: the case of Abell 3667"}, Monthly Notices of the Royal Astronomical Society Volume 445, Issue 1, p.330-346 (2014)


\end{thebibliography}


\end{document}